%% file: Aerts_liege_rev.tex
\documentclass[mypaper,7pt,twoside]{CoAst}
\usepackage{epsf,graphicx,fancyhdr}
\input{CoAst_liege-proceedingslayo}

\def\rfr{\smallskip\par\noindent
        \hangindent=7truemm
        \hangafter=1}

\begin{document}
\sf

\chapterCoAst{On the origin of macroturbulence in hot stars}
{On the origin of macroturbulence in hot stars}

\Authors{C.\ Aerts$^{1,2}$, J.\ Puls$^3$, M.\ Godart$^4$, M.-A.\ Dupret$^5$}

\Address{$^1$ Instituut voor Sterrenkunde, Celestijnenlaan 200D, B-3001 Leuven,
  Belgium \\ $^2$ Department of Astrophysics, IMAPP, Radboud University
  Nijmegen, PO Box 9010, 6500 GL, Nijmegen, the Netherlands\\$^3$
  Universit\"ats-Sternwarte, Scheinerstrasse 1, D-81679 M\"unchen, Germany\\
  $^4$ Institut d'Astrophysique et G\'eophysique, Universit\'e de Li\`ege,
  all\'ee du Six Ao\^ut 17, B-4000 Li\`ege, Belgium\\ $^5$ Observatoire de
  Paris, LESIA, 5 place Jules Janssen, 92195 Meudon Principal Cedex, France}

\noindent
\begin{abstract}
Since the use of high-resolution high signal-to-noise spectroscopy in the study
of massive stars, it became clear that an ad-hoc velocity field at the stellar
surface, termed macroturbulence, is needed to bring the observed shape of
spectral lines into agreement with observations. We seek a physical explanation
of this unknown broadening mechanism. We interprete the missing line broadening
in terms of collective pulsational velocity broadening due to non-radial
gravity-mode oscillations. We also point out that the rotational velocity can be
seriously underestimated whenever the line profiles are fitted assuming a
Gaussian macroturbulent velocity rather than an appropriate
pulsational velocity expression.
\end{abstract}

\Session{ \one } 

\section*{Macroturbulence in hot massive stars}

Velocity fields of very different scales occur in the atmospheres of hot massive
stars. Apart from the rotational velocity, which is usually assumed to be
uniform across the stellar disk and which can vary from zero speed up to the
critical value (of several hundred km\,s$^{-1}$), line-prediction codes also
include a certain amount of microturbulence (of order a few km\,s$^{-1}$) to
bring the observed profiles into agreement with the data. Microturbulence is
related to velocity fields with a scale that is shorter than the mean free path
of the photons in the atmosphere (we refer to \textit{Cantiello, these
proceedings}, for a thorough explanation).

In recent years, the number of hot massive stars that have been studied with
high-resolution spectroscopy with the goal to derive high-precision fundamental
parameters has increased quite dramatically (e.g,
\textit{Ryans et al.\ 2002, Lefever et al.\ 2007, Markova \& Puls 2008} and
references therein). This has led to the need to introduce an ad-hoc velocity
field, termed macroturbulence, at the stellar surface in order to explain the
high-quality data to an appropriate level.  This need for macroturbulent
broadening was, in fact, already emphasized by \textit{Howarth et al.\ (1997)}
from his study of massive stars from low-resolution UV spectroscopy from the
space mission IUE.  In contrast to microturbulence, macroturbulence refers to
velocity fields with a scale longer than the mean free path of the photons. In
practice, the studies listed above resulted in the requirement to introduce
macroturbulence of the 
order of several tens of km\,s$^{-1}$, and quite often even
supersonic velocity fields.

The abovementioned studies including macroturbulence rely on single snapshot
spectra and did not take 
into account pulsational velocity fields so far, as it is
done in time-resolved high-resolution spectroscopy of pulsating hot stars (e.g.,
\textit{Aerts \& De Cat 2003}).  A natural step is to investigate whether the
needed macroturbulence may simply be due to the omission of pulsational
broadening in the line-prediction codes used for fundamental parameter
estimation. In fact, for pulsating stars along the main sequence, one also needs
to add some level of macroturbulence whenever one ignores (some of) the detected
pulsations in line profile fitting of time-resolved or averaged spectra (e.g.,
\textit{Aerts \& De Cat 2003; Morel et al.\ 2006}). We investigate this
hypothesis in the present paper.

\section*{Line-profile computations in the presence of stellar oscillations}

\figureDSSN{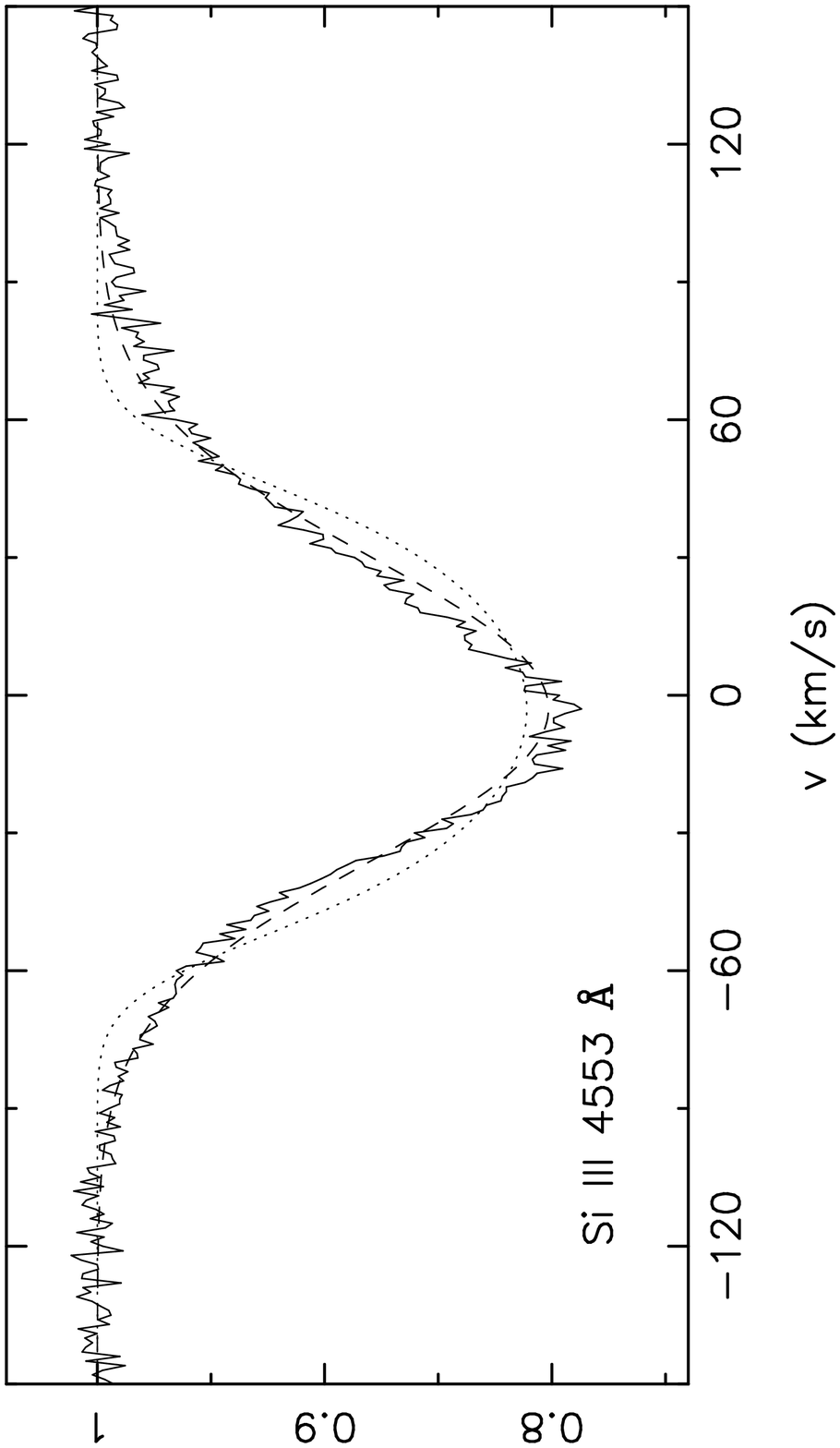}{A simulated Si line in the spectrum of a
massive hot star (full line) including, besides microturbulence and rotation,
also the collective effect of numerous very low-amplitude gravity-mode
oscillations which broaden the line wings. The input rotation was $v\sin\,i
=45\,$km\,s$^{-1}$. This profile is fitted with a model taking only rotational
broadening and microturbulence into account (dotted line, estimated 
$v\sin\,i$ 
is 57\,km\,s$^{-1}$) as well as with a model including microturbulence,
rotation and Gaussian macroturbulence (dashed line, estimated 
$v\sin\,i$ and macroturbulence are, respectively, 5 and 32\,km\,s$^{-1}$).}
{snapshot}{!t}{clip,angle=270,width=110mm}

\figureDSSN{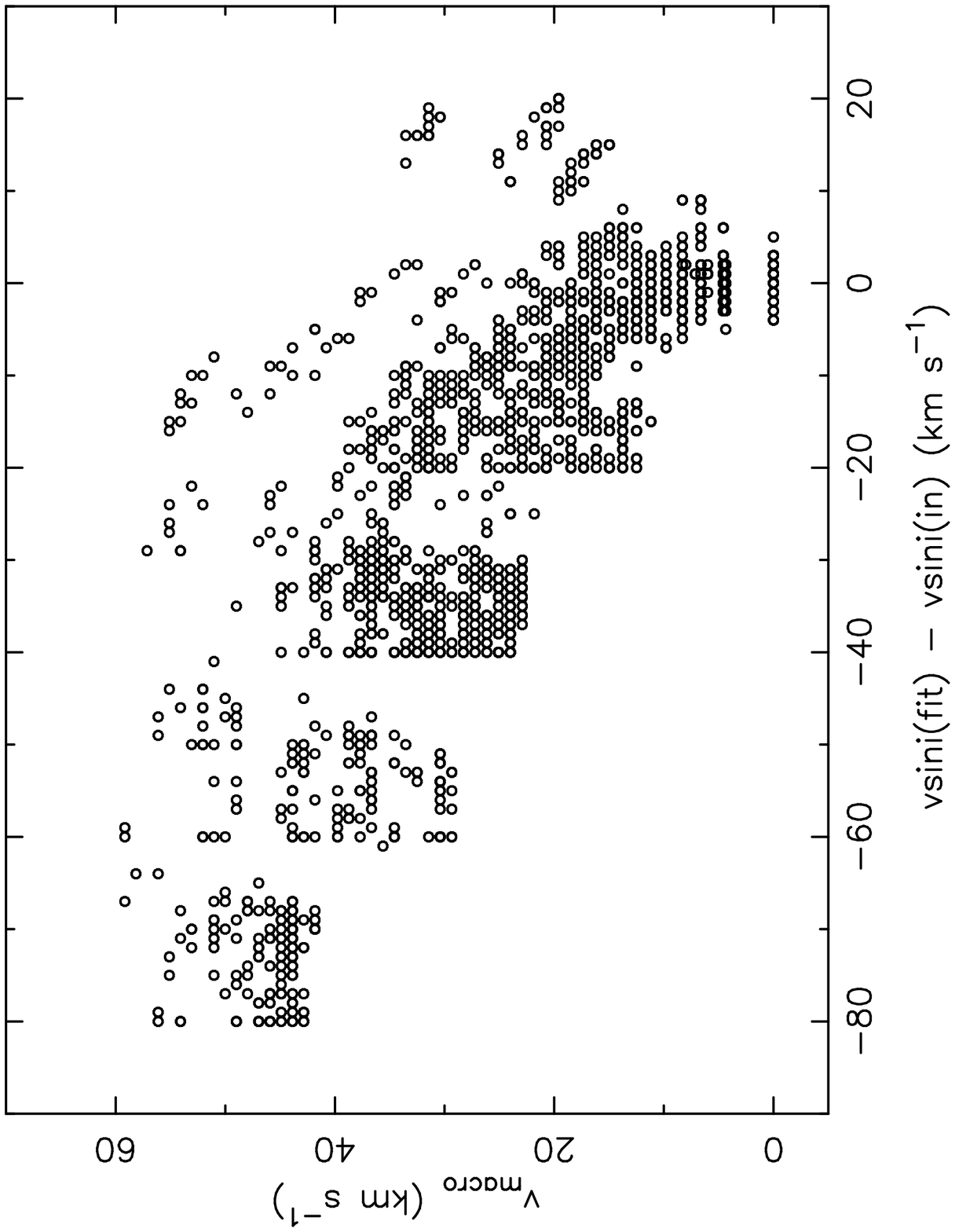}{The 
macroturbulent velocity as a function of
the rotational velocity estimate, derived from line profile fits ignoring the
presence of pulsational broadening but allowing an ad-hoc Gaussian
macroturbulence parameter, for the simulations described in the
text.}{vmacro}{!t}{clip,angle=270,width=110mm}

\figureDSSN{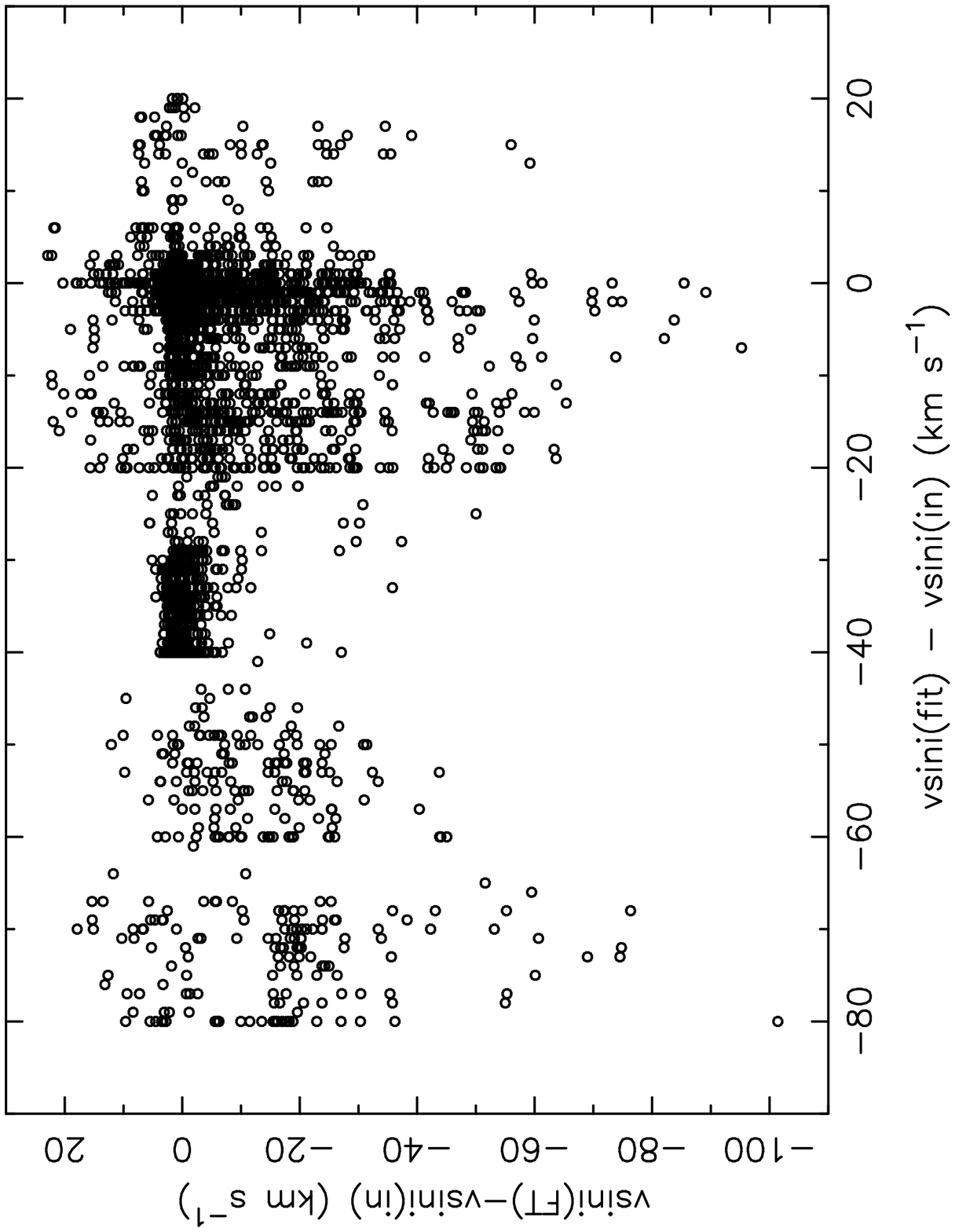}{The excess of the rotational velocity
estimated from the Fourier transform method (FT) versus the input value is
plotted as a function of the excess of the estimated rotational velocity from
line profile fits allowing for macroturbulence (fit) versus the input
value.}{vrot}{!t}{clip,angle=270,width=110mm}

We computed the excited non-radial oscillations with mode degree up to 10 for a
stellar evolution model representative of the pulsating evolved B1Ib star
HD\,163899 with the Code Li\'egeois d'\'Evolution Stellaire (\textit{Scuflaire
et al.\ 2008}) and with the non-adiabatic pulsation code MAD (\textit{Dupret
2001}).  This model has $T_{\rm eff}=18,200$\,K, $\log\,g=3.05$,
$R/R_\odot=17.8$, $\log(L/L_\odot)=4.5$, $M/M_\odot=13$, $Z=0.02$ and an age of
13 million years. The 241 excited $m=0$ modes are all gravity modes with
frequencies between 0.08 to 0.68 cycles per day and ratios of horizontal to
vertical velocity displacement in the range 0.3 to 25. They give rise to 2965
rotationally split components. These were used to compute 504 sets of
time-resolved spectroscopic line profiles of 50 profiles each, with peak-to-peak
amplitudes for the radial velocity between 0.7 to 15\,km\,s$^{-1}$ and for
rotational velocities between 25 and 125\,km\,s$^{-1}$, which is 
below 25\% of the critical value, following the method of
\textit{Aerts et al.\ (1992)}.  Such radial-velocity amplitudes are well below
those observed for hot B and A supergiants (e.g., \textit{Kaufer et al.\ 1997,
Prinja et al.\ 2004}) so that we can be sure not to have overestimated the
effects of oscillations on the lines. These profiles were then fitted ignoring
the oscillations but allowing for a Gaussian macroturbulent velocity.

{Using a goodness-of-fit approach,}
we confirm the finding that the inclusion of an ad-hoc macroturbulence parameter
leads to better fits than those obtained when only allowing rotational and
microturbulent broadening (Fig.\,\ref{snapshot}). The pulsational broadening
ignored in the line profile fits is compensated by allowing a macroturbulent
velocity. These ad-hoc velocities are sometimes in excess of the speed of sound
to ensure a good fit 
(Fig.\,\ref{vmacro}). At first sight, it might seem surprising to need such high
macroturbulent speeds to explain the collective effect of oscillation modes
that have by themselves only very low velocity amplitude. However, this is easy
to understand if one realises that line widths depend on velocity squared, such
that numerous small velocities add up to give a very significant effect in the
overall line broadening when the collective effect of the modes is interpreted
by a single ad-hoc parameter. This is particularly the case for gravity-mode
oscillations which impact strongly on line wings. On the other hand, such
low-amplitude modes do not alter seriously the observed quantities behaving
linearly with velocity, such
as the radial velocity variations of the star, 
because their collective effect tends to cancel
out in this case.

As a side result of our line fits with macroturbulence, we report a risk to
underestimate the projected rotational velocity appreciably when using the
Fourier Transform (FT) method 
to estimate the projected rotational velocity from
the first minimum of the FT (Sim{\'o}n-D{\'{\i}}az \& Herrero 2007), as is
illustrated in Fig.\,\ref{vrot}. While this method works well in general and is
able to recover the correct input value of the rotational broadening for most of
the cases, it is sometimes fooled when too asymmetric pulsational broadening is
present and, in this case, one derives too low estimates for $v\sin\,i$. 
{This is also the case for the results from a goodness-of-fit approach}
(see Fig.\,\ref{snapshot}).

\section*{Conclusions}

We have shown that pulsational velocity broadening due to the collective effect
of numerous low-amplitude gravity mode oscillations offers a natural and
appropriate physical explanation for the occurrence of macroturbulence in hot
massive stars. Our computations of course do not exclude other and/or additional
physical interpretations of the macroturbulent velocities reported in the
literature. 

We ignored rotational and non-adiabatic effects in the computations of the
velocity eigenfunctions of the non-radial modes used for the line profile
simulations. Codes to incorporate each of these effects separately are available
(e.g., \textit{Townsend 1997, De Ridder et al.\ 2002}) and their influence on
the line profiles are well understood. In particular, they will not alter the
line wings of the profiles dramatically as long as the rotational velocity
remains below 25\% of the critical velocity (\textit{Aerts \& Waelkens 1993})
and would thus not alter the main conclusions of our work, while implying a very
serious increase in CPU time.

From the observational side, high-precision multi-epoch observations of
unblended metal lines are needed to evaluate appropriately the effect of
pulsational broadening.  Such type of data have not been used so far to estimate
macroturbulence.

\acknowledgments{The computations for this research have been done on the VIC
HPC supercomputer of the K.U.Leuven.  CA is supported by the Research Council of
K.U.Leuven under grant GOA/2008/04.  }

\References{

\rfr Aerts C., \& De Cat P.\ 2003, SSRv, 105, 453

\rfr Aerts C., De Pauw M., \& Waelkens C.\ 1992, A\&A, 266, 294

\rfr Aerts C., \& Waelkens C.\ 1993, A\&A, 273, 135

\rfr De Ridder J., Dupret M.-A., Neuforge C., \& Aerts C.\ 2002, A\&A, 385, 572

\rfr Dupret M.-A.\ 2001, A\&A, 366, 166

\rfr Howarth I.~D., Siebert K.~W., Hussain G.~A.~J., \& Prinja R.~K.\ 1997,
  MNRAS, 284, 265

\rfr Kaufer A., Stahl O., Wolf B., et al.\ 1997, A\&A, 320, 273

\rfr Lefever K., Puls J., \& Aerts C., 2007, A\&A, 463, 1093

\rfr Markova N., \& Puls J.\ 2008, A\&A, 478, 823

\rfr Morel T., Butler K., Aerts C., et al.\ 2006, A\&A, 457, 651

\rfr Prinja R.~K., Rivinius Th., Stahl O., et al.\ 2004, A\&A, 418, 727

\rfr Ryans R.~S., Dufton P.~L., Rolleston W.~R.~J., et al.\ 2002,
MNRAS, 336, 577

\rfr Scuflaire R., Th\'eado S., Montalb\'an J., et al.\ 2008, ApSS, in press

\rfr Sim{\'o}n-D{\'{\i}}az S., \& Herrero A.\ 2007, A\&A, 468, 1063

\rfr Townsend R.~H.~D.\ 1997, MNRAS, 284, 839

}
\end{document}

%% file: CoAst_liege-proceedingslayo.tex
\renewcommand{\chaptermark}[1]{\markboth{#1}{}}

\newcommand{\kopf}{\small\itshape Comm. in Asteroseismology \\ Contribution to the Proceedings of the 38$^{th}$\,LIAC\,/\,HELAS-ESTA\,/\,BAG, 2008
}

\newcommand{\Authors}[1]{\begin{center}\normalsize\bf\sf #1 \end{center}}

\renewcommand{\author}[1]{\begin{center}\normalsize\bf\sf #1 \end{center}}
\newcommand{\Address}[1]{\begin{center}\small\sf #1 \end{center}}

\DeclareGraphicsExtensions{.eps,.jpg}

\newcommand{\Session}[1]{{\vspace{3mm}\small \noindent  \hspace*{3mm} Session: } #1 \normalsize}

	\newcommand{\one}{\small Physics and uncertainties in massive stars on the MS and close to it}

\renewenvironment{abstract}{\section*{Abstract}\normalsize\sf}{}
\newcommand{\References}[1]{\begin{flushleft}{\large References\\}\vspace*{2mm}\small #1 \end{flushleft}}

\newcommand{\chapterCoAst}[2]{\chapter[\sf\normalsize #1\\ \footnotesize \hspace*{5mm}by #2 \sf\normalsize][]{#1\\}\rhead[\fancyplain{}{\sf\footnotesize \center{#1}}]{\fancyplain{}{\sffamily\thepage}}\lhead[\fancyplain{\kopf}{\sffamily\thepage}]{\fancyplain{\kopf}{\sf\footnotesize \center{#2}}}}

\newcommand{\figureDSSN}[5]{\begin{figure}[#4]
\centering
\includegraphics*[#5]{#1}
\caption{#2}
\label{#3}
\end{figure}}

\renewcommand{\chaptername}{}

\newcommand{\acknowledgments}[1]{\vspace*{5mm}\noindent  \textbf{Acknowledgments.} #1}